\documentclass[]{spie}  %>>> use for US letter paper
\usepackage{amsmath,amsfonts,amssymb}
\usepackage{graphicx}
\usepackage[colorlinks=true, allcolors=blue]{hyperref}\usepackage{xspace}
\usepackage{upgreek}
\usepackage{amsmath,amsfonts,amssymb}
\usepackage{graphicx}
\usepackage{color, colortbl} %need for color tables like from Excel2Latex
\usepackage{booktabs} %for making fancy lines in tables like Excel2Latex
\usepackage{multirow} %for making multirow table entries
\usepackage[inline]{enumitem} % for enumerate*, an inline enumerate environment

\usepackage{wasysym}
\usepackage{physics} %has nice derivative expressions \dv{}{} and \pdv{}{} and \dd{}
\usepackage{textcomp,gensymb} % has the \degree command, if you don't also load text comp it gives some weird warnings about not defining \perthousand and \micro
\usepackage{subcaption} %for subfigures
\usepackage{siunitx}

\usepackage{etoolbox,pdftexcmds}
\makeatletter
\patchcmd{\footnote}
  {\stepcounter\@mpfn}
  {\stepcounter\@mpfn\check@overflow\@mpfn}
  {}{}
\newcommand{\check@overflow}{%
  \ifnum\pdf@strcmp{\@mpfn}{footnote}=\z@
    \ifnum\value{footnote}>9
      \setcounter{footnote}{1}%
    \fi
  \fi
}
\makeatother

\renewcommand{\deg}{$^{\circ}$\xspace }

\newcommand{\um}{$\upmu$m\xspace}	%xspace makes it not leave a weird space

\def\arcmin{\hbox{$^\prime$}\xspace} %arcminute
\def\arcsec{\hbox{$^{\prime\prime}$}\xspace} %arcsecond

\title{In-flight performance of the BLAST-TNG telescope platform}

\author[a,b]{Gabriele Coppi}
\author[c]{Peter A.R. Ade}
\author[d]{Peter C. Ashton}
\author[e]{Jason E. Austermann}
\author[k]{Erin G. Cox}
\author[b]{Mark J. Devlin}
\author[e]{Bradley J. Dober}
\author[a]{Valentina Fanfani}
\author[f]{Laura M. Fissel}
\author[g]{Nicholas Galitzki}
\author[e]{Jiansong Gao}
\author[h]{Samuel Gordon}
\author[h]{Christopher E. Groppi}
\author[e]{Gene C. Hilton}
\author[e]{Johannes Hubmayr}
\author[b]{Jeffrey Klein}
\author[i]{Dale Li}
\author[j]{Nathan P. Lourie}
\author[b]{Ian Lowe}
\author[h]{Hamdi Mani}
\author[h]{Philip Mauskopf}
\author[e]{Christopher McKenney}
\author[a]{Federico Nati}
\author[k]{Giles Novak}
\author[c]{Giampaolo Pisano}
\author[l]{L. Javier Romualdez}
\author[h]{Adrian Sinclair}
\author[n]{Juan D. Soler}
\author[c]{Carole Tucker}
\author[e]{Joel Ullom}
\author[e]{Michael Vissers}
\author[m]{Caleb Wheeler}
\author[k]{Paul A. Williams}

\affil[a]{University of Milano-Bicocca, Piazza della Scienza 3, 20126 Milano (MI), Italy}
\affil[b]{University of Pennsylvania, 209 South 33rd Street, Philadelphia, PA 19104, USA}
\affil[c]{Cardiff University, The Parade, Cardiff CF24 3AA, United Kingdom}
\affil[d]{Lawrence Berkeley National Laboratory, 1 Cyclotron Rd, Berkeley, CA 94720}
\affil[e]{NIST-Boulder, 325 Broadway, Boulder, CO 80305}
\affil[f]{Queen's University, 99 University Ave, Kingston, ON K7L 3N6, Canada}
\affil[g]{University of California San Diego, 9500 Gilman Dr, La Jolla, CA 92093}
\affil[h]{Arizona State University, 550 E Tyler Drive, Tempe, AZ 85287}
\affil[i]{SLAC National Accelerator Laboratory, 2575 Sand Hill Rd, Menlo Park, CA 94025}
\affil[j]{MIT Kavli Institute for Astrophysics and Space Research, 70 Vassar St, Cambridge, MA 02139}
\affil[k]{Northwestern University, 1800 Sherman Ave, Evanston, IL 60201}
\affil[l]{Princeton University, Jadwin Hall, Washington Road, Princeton, NJ 08544}
\affil[m]{Underground Instruments, 224 Claremont Ave. San Antonio TX 78209}
\affil[n]{Max-Planck-Institut für Astronomie, Königstuhl 17
69117 Heidelberg, Germany}

\authorinfo{Further author information: (Send correspondence to Gabriele Coppi)\\Gabriele Coppi: E-mail: gabriele.coppi@unimib.it}

\begin{document} 
\maketitle

\begin{abstract}
    The Next Generation Balloon-Borne Large Aperture Submillimeter Telescope (BLAST-TNG) was a unique instrument for characterizing the polarized submillimeter sky at high-angular resolution. BLAST-TNG flew from the Long Duration Balloon Facility in Antarctica in January 2020. Despite the short flight duration, the instrument worked very well and is providing significant information about each subsystem that will be invaluable for future balloon missions. In this contribution, we discuss the performance of telescope and gondola. 
\end{abstract}

\section{Introduction}
\label{section:intro}
The Next Generation Balloon-Borne Large Aperture Submillimeter Telescope (BLAST-TNG) was a unique instrument for characterizing the polarized submillimeter sky at high-angular resolution. BLAST-TNG was designed to study the role of magnetic fields in shaping the structure and evolution of the interstellar medium (ISM) and the role of magnetic fields in regulating star formation by mapping polarized dust emission. BLAST-TNG was designed to have high angular resolution and sensitivity to complement  current and planned submillimeter observatories like \textit{Planck} \cite{planck_XIX}, ALMA \cite{alma}, \textit{SOFIA} \cite{sofia}, POL-2 on the James Clark Maxwell Telescope \cite{pol2_jcmt}, and CCAT-Prime \cite{ccat_prime_parshley}. 

BLAST-TNG featured three microwave kinetic inductance detector (MKID) arrays operating over 30\% bandwidths centered at 250, 350, and 500 \um. In the three bands, these highly-multiplexed, high-sensitivity arrays include 918, 469, and 272 dual-polarization pixels respectively, for a total of more than 3,000 detectors. The detectors were coupled to a cassegrain telescope through a cold reimaging optics system resulting in a diffraction-limited instrument with a resolution of 30, 41 and 59 arcsec. Simultaneous observations were achieved through the use of two dichroic beam splitters. The arrays are cooled to $\sim$275 mK in a liquid-helium-cooled cryogenic receiver, with a 250 L reservoir which can operate for \textgreater 24 days. BLAST-TNG was launched on January 6th, 2020 and made observations during an 18 hour flight from McMurdo Station in Antarctica as part of NASA's long-duration-balloon (LDB) program. Despite the short flight, the team was able to get valuable data about the performance of the detectors and other subsystems that would prove their functionality on a balloon platform. Much of the instrument (including the receiver) was destroyed due to a rough impact and drag upon landing, however the primary mirror and many subsystems (such as the readout and control electronics, motors, and hard drives) were successfully recovered during an intense recovery campaign.

\begin{figure*}
    \centering
    \includegraphics[height=4in]{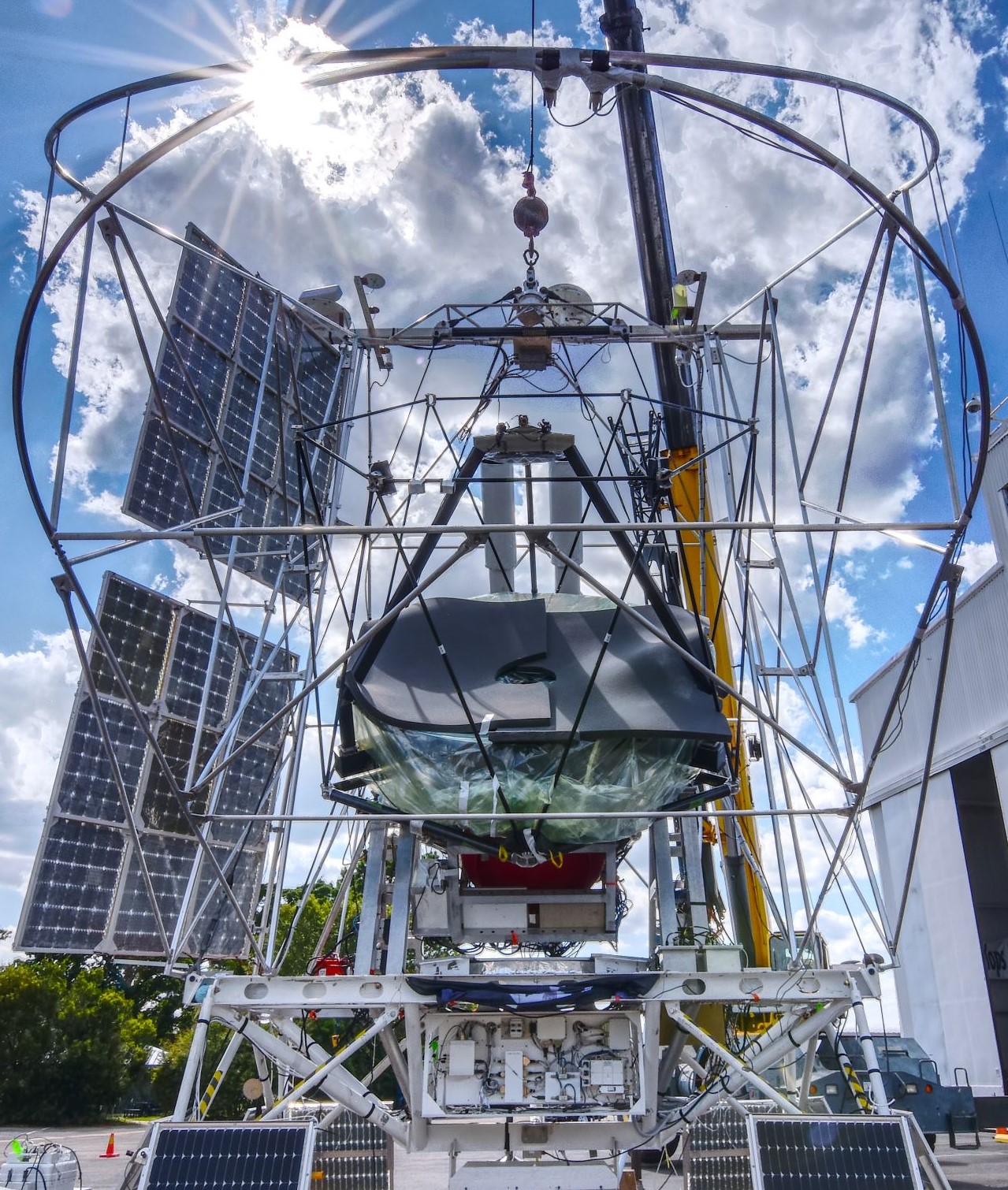}
    \hfill
    \includegraphics[height=4in]{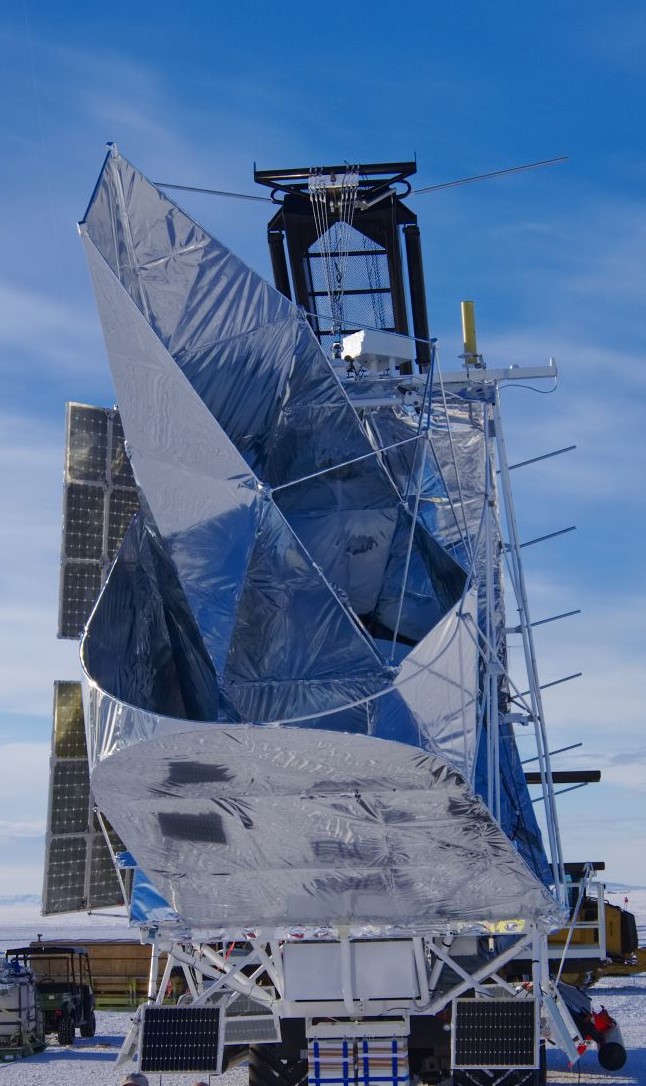}
    \caption{Photos of the fully assembled BLAST payload in Palestine, TX with no sunshields and in Antarctica with sunshields.}
    \label{fig:BLASTPhotos}
\end{figure*}

The BLAST-TNG observations were designed to fulfill several scientific goals, including to: 
\begin{enumerate*}[label=(\arabic*)] 
\item obtain a sample of deep maps of star-forming regions with sizes ranging from 0.25 to 20 square degrees to better understand how magnetic fields influence the formation and evolution of molecular clouds;
\item apply statistical methods to quantify the observed magnetic field morphology. By measuring the magnetic field, its dispersion, and their correlations with other physical parameters of the ISM, constraints can be put on field strength, the turbulent power spectrum, and models of cloud formation \cite{houde_2009_turbulence}$^,$\cite{soler_hennebelle}$^,$\cite{soler_sims_2013}$^,$\cite{jow_hro}$^,$\cite{fissel_blastpol} ;
\item map magnetic fields in a large sample of dense prestellar and protostellar cores, and show how the core magnetic field relates to the larger scale cloud fields; 
\item map diffuse interstellar dust to constrain models of dust composition and alignment mechanisms through measurements of changes in polarization as a function of frequency \cite{draine_fraisse_2009,draine_hensley} ;
\item characterize the (polarized) diffuse dust emission on small angular scales; \item serve as a true ``observatory'' platform on which science groups outside of the BLAST-TNG collaboration can propose observations, up to 25$\%$ of the observation time was reserved for the scientific community.\end{enumerate*} In this paper we describe the pointing system and its in flight performance in section \ref{section:pointing}. Following this section we discuss the inflight behaviour of two key subsystems for a balloon experiment, the power system, section \ref{sec:power}, and the thermal behaiour of the system, section \ref{sec:thermal}. Finally, we discuss the optical performance in section \ref{sec:optics}.

\section{Overview}
\label{section:payload}

\begin{figure}
    \centering
    \includegraphics{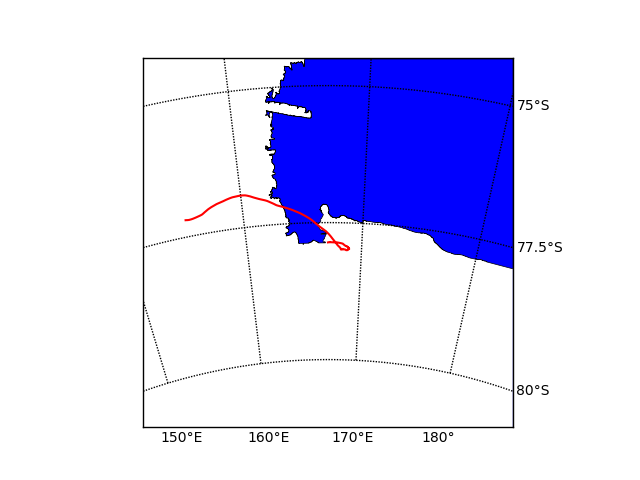}
    \caption{Flight path of the BLAST-TNG payload}
    \label{fig:flight}
\end{figure}

BLAST-TNG was a particularly complex payload with multiple subsystems that work together to allow the observations. In this paper we focus on all the subsystems except the cryogenics receiver and detectors which are described in a companion paper \cite{Lowe2020}. The fully assembled payload is shown in figure \ref{fig:BLASTPhotos}. Here, the payload is presented during the compatibility test in Palestine, TX at the CSBF facility with no sunshields to show all the components. There is also a photo of the payload in Antarctica just before a launch opportunity.
BLAST-TNG flew from the Antarctic LDB facility in January 2020. The trajectory of the flight is presented in figure \ref{fig:flight}. Unfortunately, during launch operations, the collar hit the payload (every launch has the intrinsic risk of the payload being hit by the collar, but statistics of previous launches show that the probability is very low) and in particular one of the structural components, the spreader bar. The damage was not immediately spotted since this component was designed with a very high factor of safety. However, the carbon fiber structure slowly degraded until a catastrophic failure after approximately 14 hours of flight. Despite this unfortunate event, every subsystem worked as expected as will be described in the following sections with some preliminary analysis. 

\section{Pointing}
\label{section:pointing}

BLAST-TNG is a scanning telescope, and the axes we pointed at are azimuth (az) and elevation (el). While observing a source, we generally scanned quickly (0.1\deg $s^{-1}$ - 0.2\deg $s^{-1}$) in az while slowly stepping in el. To control the attitude of the telescope, we had three DC motors, the reaction wheel motor, the pivot, and the elevation drive. The first two control the movement of the the telescope in azimuth, while the latter takes care of changing the elevation of the telescope. However, the motors are only a part of our global pointing system. During each scan we need to reconstruct the position and to achieve this result we combine multiple sensors as described in subsection \ref{sec:pointing sensors}. Finally, the software that controls both motors and computes the in-flight pointing solution is described in subsection \ref{sec:pointing control}.

\subsection{Pointing Sensors}
\label{sec:pointing sensors}

Our attitude determination system consists of a combination of multiple sensors that can be divided into three categories: high-accuracy ``fine'' pointing sensors, low-accuracy ``coarse'' pointing sensors, and gyroscopes. The sensors that belong to the first two categories are combined together weighted by each sensor accuracy. The time dependent pointing solution is created by combining also the data coming from the gyroscopes. 

Our high-accuracy pointing sensors were two identical, optical star cameras (properties in Table \ref{tab:sc}), and a high resolution elevation encoder. Both star cameras were triggered to take images at the turnarounds in azimuth, where telescope's angular velocity is at its minimum and the stars in the images would not be streaked for our typical integration time. 
Such strategy allowed to take images every ~20 seconds approximately, depending on the scan size and speed. 

\begin{table*}
     \centering
     \begin{tabular}{rc}
          \toprule
          \multicolumn{2}{c}{Star Camera Properties}\\
          \midrule
          Pixels & 1392 $\times$ 1040\\
          Pixel Size & 6.45 \um $\times$ 6.45 \um\\
          Peak Quantum Efficiency & 60\%\\
          Dynamic Range & 14 bit\\
          Well Depth & 16,000 e$^-$\\
          Read Noise & 6.5 e$^-$\\
          Pixel FOV & $7\arcsec$\\
          Camera FOV & 2.5\deg $\times$ 2\deg\\
          Lens Diameter & 100 mm\\
          Lens F/\# & 2 \\
          Lens Optical Efficiency & 0.8 \\
          Filter cut-off & 600 nm\\
          \bottomrule
     \end{tabular}
     \caption{Star Camera Properties: The two star cameras flown on BLAST-TNG were identical and were both mounted parallel to the boresight of the telescope.}
     \label{tab:sc}
 \end{table*}
 
The elevation encoder was read-out continuously throughout the scan, and is the only elevation sensor apart from the star cameras. However, the star cameras do not measure directly the elevation, but the celestial coordinates and then convert them to telescope coordinates. This means that the elevation encoder is our single direct measurement of the elevation. The star cameras run STARS\cite{chapman_stars}, a ``lost in space'' algorithm on every image, which does not take into account any prior pointing information in solving for a pointing solution.

During the 2020 flight of BLAST-TNG, the star cameras' performance was affected by the presence of polar mesospheric clouds (PMCs). As a result of this, on multiple images the star cameras were not able to solve the image and provide a position on the sky. However, a preliminary post-flight analysis of the images of the star-cameras increases significantly the number of pointing solution thanks to the use of advanced filtering algorithms that was not possible in flight due to the low computational power available. 

\begin{figure}
    \begin{subfigure}[t]{0.5\textwidth}
      \centering
      \includegraphics[width=\linewidth]{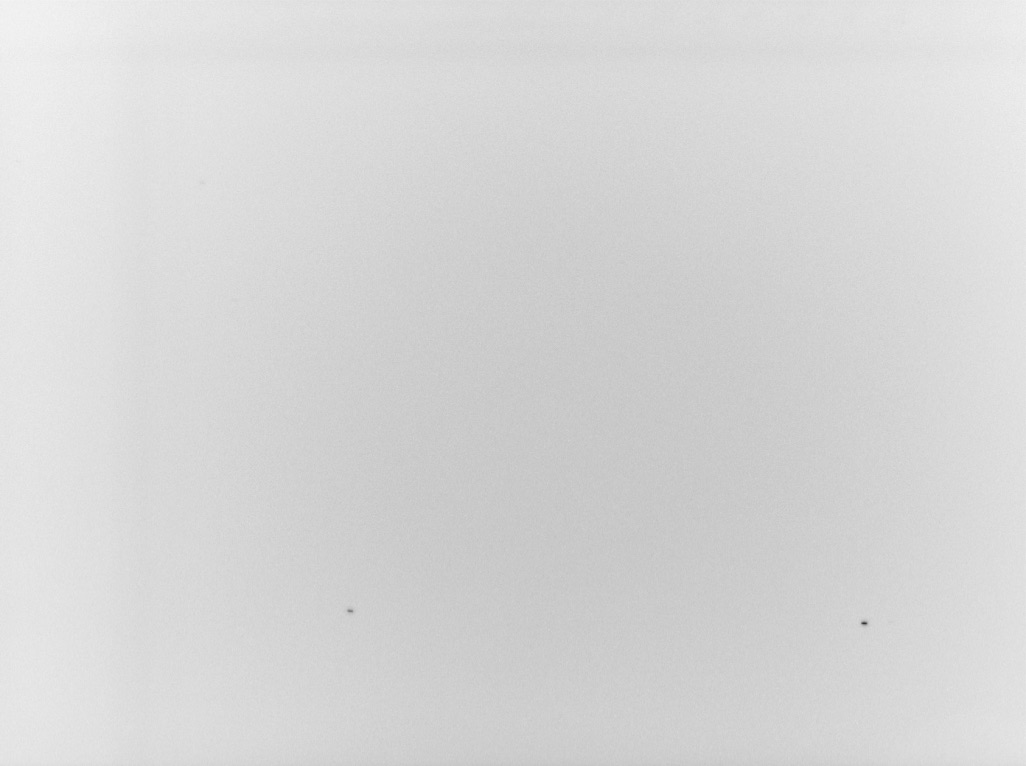}
      \caption{No PMCs}
      \label{fig:xsc_no_pmc}
    \end{subfigure}
    \hfill
    \begin{subfigure}[t]{0.5\textwidth}
      \centering
      \includegraphics[width=\linewidth]{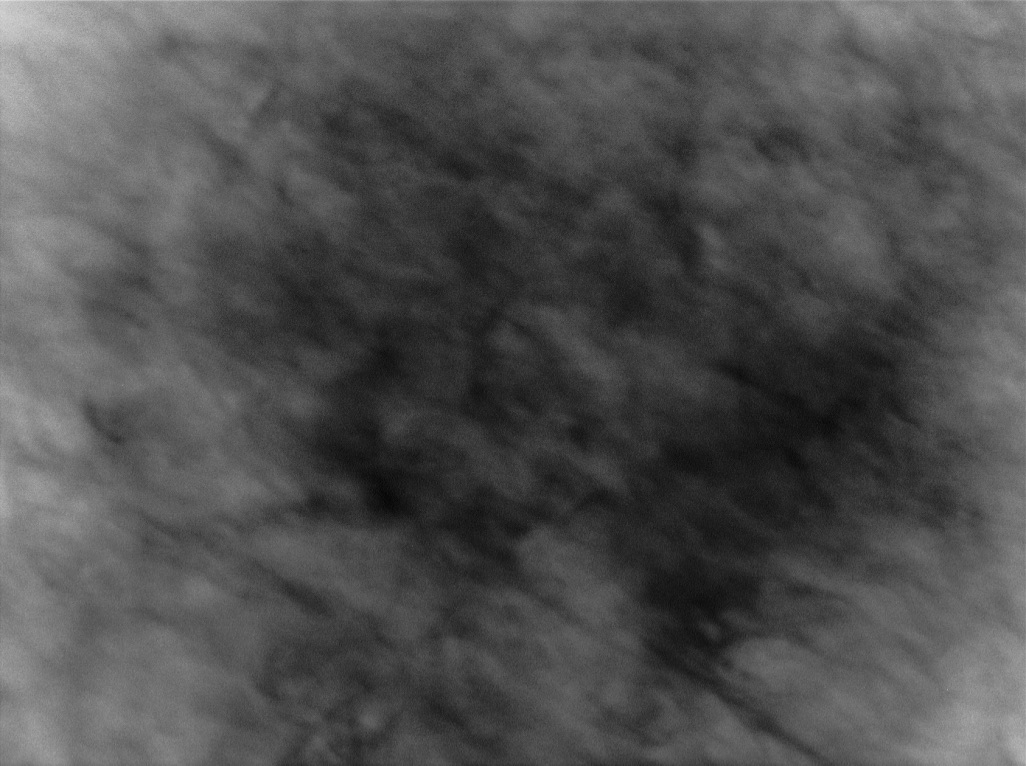}
      \caption{Full coverage by PMCs}
      \label{fig:xsc_pmc}
    \end{subfigure}
    \caption{Example of a clean star camera image and one contaminated by PMCs.}
\end{figure}

\begin{table*}
     \centering
     \begin{tabular}{lS}
        \toprule
          Type of Solution & \multicolumn{1}{c}{Number of Images Solved}\\
          \midrule
            During Flight & 593\\
            Total post-flight & 1983\\
            $\qquad$Post-flight, no filtering & 333 \\
            $\qquad$Post-flight, with filtering & 1650\\
          \bottomrule
     \end{tabular}
     \caption{Star Camera Performance: The total images solved during and after the flight, using STARS\cite{chapman_stars} during the flight and software from astrometry.net\cite{astrometry_net} post-flight.}
     \label{tab:sc solutions}
 \end{table*}

The low-accuracy or ``coarse'' pointing sensors serve two purposes. The first is to give rough pointing solutions when the star cameras are not able to get solutions. Additionally, the coarse pointing sensors are used to correct for noise in the gyroscope signal. The gyroscopes (KVH Industries DSP-1760) present a low frequency noise componenent that manifests as a slowly varying DC offset between the true and measured velocities. Each coarse pointing sensor report the true velocity and this value is compared to the velocity measured by the gyroscopes to estitmate the DC offset.
Our coarse pointing sensors include pinhole sun sensors, differential GPS (provided by CSBF and used as an azimuth sensor), magnetometers, and inclinometers, and they are summarized in Table \ref{tab:sensors}.

 \begin{table}
     \centering
     \begin{tabular}{cccc}
     \toprule
        Sensor & Location & Rate(Hz) & Accuracy(\deg) \\
        \midrule
        Star Cameras & Inner Frame & 0.05-0.1 & \textless0.001\\
        Elevation Encoder & Outer Frame & 100 &\textless0.01\\
        Pinhole Sun Sensors & Outer Frame & 5 & 0.2 \\
        Magnetometers & Outer Frame & 100 &5\\
        DGPS & Outer Frame & 5 &0.2\\
        Inclinometers & Both Frames & 5& 0.1\\
        \bottomrule
     \end{tabular}
     \caption{Pointing Sensor Properties}
     \label{tab:sensors}
 \end{table}
 
The pinhole sun sensors used Hamamatsu S5991-01 position sensitive diodes (PSDs). When the spot of light from the 200 \um pinhole light strikes the PSD, charge proportional to the intensity of the light is generated. When voltage biased, current flows to the four electrodes at the corners of the active area, and the relative magnitudes of these currents can be used to determine the spot location. Initial results from the pinhole sun sensors analysis from the 2020 BLAST-TNG flight show a pointing precision of 0.2\deg.

We mounted two magnetometers on the sun shields, as far as possible from any active components to minimize the effect of magnetic fields induced by current in this component. As previously mentioned, magnetometers, as with the rest of the coarse sensors, provide information about the azimuth. However, due to the nature of these sensors, their use in a polar flight is particularly challenging. These sensors measure the horizontal component of the Earth magnetic field to get the azimuth solution. However, close to the geomagnetic pole the magnetic field has almost only a vertical component and a very weak horizontal component. 

Finally, we also flew two capacitive inclinometers, that were mainly useful to reconstruct the event that caused the end of our flight. They confirmed that the outer frame of the gondola was pitched forward by 4\deg, which was consistent with comparisons between the star camera elevation solutions and the elevation encoder.

\subsection{Pointing Control}
\label{sec:pointing control}
Our pointing control software uses a proportional-integral-derivative (PID) control loop.
For all our scans, we set a desired scan speed, or \emph{set-point}. The PID loop first calculates the error between the measured scan speed and the set-point. This error is represented by $e(t)$, and is a function of time. To find the motor current that we command at the next time step, we calculate equation \ref{eqn:pid}, where $i(t)$ is the motor current. $K_\text{p}, K_\text{i}$ and $K_\text{d}$ are coefficients that adjust the relative strengths of the proportional, integral, and derivative terms, respectively. Ground and flight tests showed that the derivative term did not significantly improve our pointing performance, so to simplify operations it was set to zero. 

\begin{equation}
\label{eqn:pid}
    i(t) = K_\text{p} e(t) + K_\text{i} \int_0^t e(t') \,dt' + K_\text{d} \frac{d}{dt}e(t)    
\end{equation}

The elevation control loop is relatively simple, it simply follows the equation \ref{eqn:pid} with specific set-point chosen based on the scan.

The azimuth control is a little bit more complicated. As previously mentioned, BLAST-TNG uses two different motors, the pivot and the reaction wheel to control the azimuth, and each one has slightly different tasks.
The reaction wheel is a large ($\sim$1 m diameter) wheel with a large moment of inertia. It is driven by a brushless DC motor that can transfer angular momentum between the reaction wheel and the rest of the payload very quickly, so by setting high PI coefficients, the reaction wheel exerts most of the torque needed for our maneuvers.
However, when the reaction wheel reaches its maximum speed of about 300\deg $s^{-1}$ or 50 rpm, its motor saturates due to back EMF. In order to avoid the reaction wheel motor reaching a saturation condition, we use the pivot to slowly shed excess angular momentum and keep the reaction wheel rotating at a speed much less than the saturation point, typically at about 30\deg $s^{-1}$. The pivot can also be used to point directly (though it does not have as much torque as the reaction wheel motor because it torques against the balloon through the flight train, which acts like a torsion spring with a very low spring constant). As result of this double role, the pivot presents two different PI terms in an equation that has the same form as the one presented in \ref{eqn:pid}.

During our 2020 flight, we used several pointing modes. The simplest mode is ``drift'', that uses feedback from the gyroscopes to move the telescope at a constant speed in both azimuth and elevation. The majority of our scans used a mode called  ``box''. In this particular mode, the center of the box is defined in RA and DEC and the box size is defined in the AZ and EL space. An extension of this mode, less used, is the ``quad'' mode. In this case, the four corners of a box are defined in RA and DEC, and then the scan happens in AZ and EL. During our observations, we find that over ten minutes, our pointing is stable to 2.4\arcsec in azimuth and 7.1\arcsec in elevation.

\section{Power Distribution}
\label{sec:power}

The BLAST-TNG payload was powered by solar panels manufactured by SBM Solar Inc. (SBM Solar Inc., 8000 Poplar Tent Rd C, Concord, NC 28027). These solar panels were tested to provide at least 100\,W of power each in optimal conditions on the ground. The inner and outer frames each draw between 400 and 500\,W during normal operation, and required additional solar capacity to power the steady state while charging the batteries during flight. To this end there were 18 panels arranged into two arrays, one for the inner frame and one for the outer frame, each one consisting of 9 panels.  For further redundancy the panels were arranged in 3 separate series of 3 panel each. Each series was producing approximately 60\,V with good illumination. 
With a 1.8\,kW charging capacity the BLAST payload was able to maintain power levels even at 50\% operating capacity. This overhead allowed for operations to continue under sub-optimal conditions, such as non-normal solar incidence, degraded performance due to high temperatures, and  launch attempts on cloudy days. During the flight the solar systems behaved nominally, providing more than enough power for the instrument during charging sessions.

The BLAST-TNG gondola is powered off of ODS-AGM42L lead acid batteries produced by Odyssey (EnerSys World Headquarters 2366 Bernville Road, Reading, PA 19605, USA), utilizing four parallel sets of two batteries. The four parallel sets were broken down into two sets of two in parallel for the inner and outer frames, providing excess capacity to lengthen observations in between charging and provide a buffer for launch operations. In order to match the 28.6\,V of the batteries, the two arrays were interfaced with a charge controller module TS-MPPT-60 manufactured by Morningstar Corporation (Morningstar Corporation, 8 Pheasant Run, Newtown, PA 18940). This system allowed a variable flow of power (varying I and V) generated by the solar panels to be converted into a maximum 28.6V potential on the batteries with immediate usage going directly to systems and excess power being used to recharge the batteries. When all the battery sets were fully charged, the charge controllers are capable of lowering the current draw on the solar panels, preventing overcharging.

\begin{figure}
    \begin{subfigure}[t]{0.5\textwidth}
      \centering
      \includegraphics[width=\linewidth]{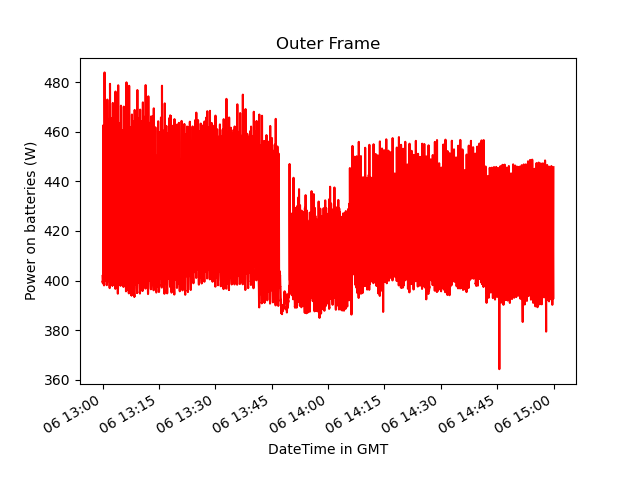}
      \caption{Outer Frame power production, battery fully charged}
      \label{fig:of}
    \end{subfigure}
    \hfill
    \begin{subfigure}[t]{0.5\textwidth}
      \centering
      \includegraphics[width=\linewidth]{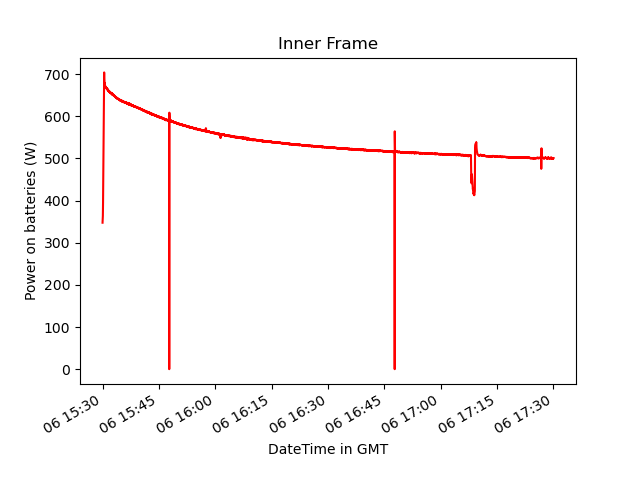}
      \caption{Inner Frame power production, from charging to battery fully charged}
      \label{fig:if}
    \end{subfigure}
    \caption{Two different examples of the the power production in flight for the two frames. The drops in the in Inner Frame plot are due to lost data packages. }
\end{figure}

The solar and battery systems acted as the source and bank for the gondola power, while the distribution network provides voltage switching, isolation, and control to the different subsystems. The power network was designed to provide independent power for each subsystem on the gondola and interfaces between the batteries and the electronics through a nexus of distribution points. Each distribution point consisted of a Vicor (Vicor Corporation 25 Frontage Road, Andover, MA 01810, USA) DC-DC converting the variable battery voltage to a steady output coupled to a Vicor $\mu RAM$ ripple attenuation module which serves to further reduce voltage variability. The distribution points were bundled together into larger distribution breakout boxes containing the power modules as well as the power connection relays controlled by the flight software via the outer frame LabJacks (LabJack Corporation 3232 S Vance St, STE 200 Lakewood, CO  80227-5030, USA) and, as a backup, the CSBF SIP science stack. The BLAST-TNG power distribution system consisted of three breakout boxes, one for the outer frame, one for the inner frame, and a special high power-draw one for the cryogenic scroll pumps. Additionally, the inner frame power was further broken down by a cryogenic readout power distribution box which provided power to the housekeeping systems in the cryostat.

\section{Thermal Performance}
\label{sec:thermal}

\begin{figure*}[htp]
    \centering
    \includegraphics[width=0.45\textwidth]{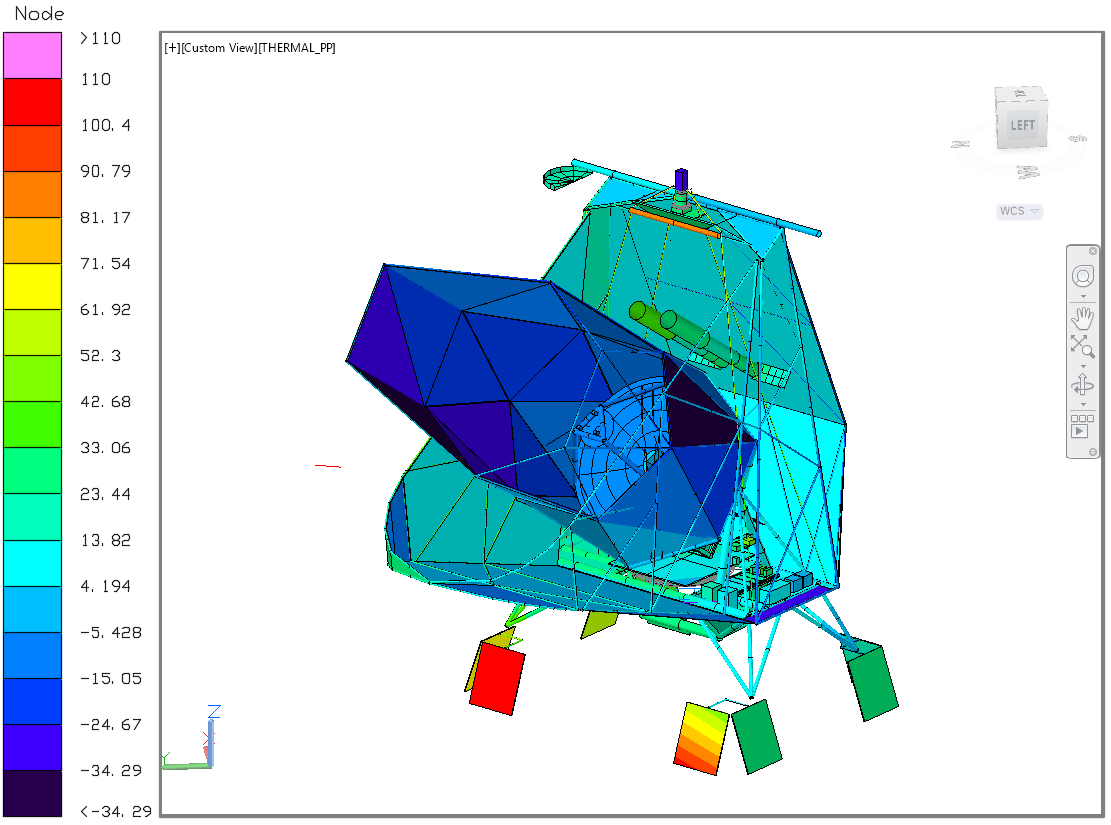}
    \hfill
    \includegraphics[width=0.45\textwidth]{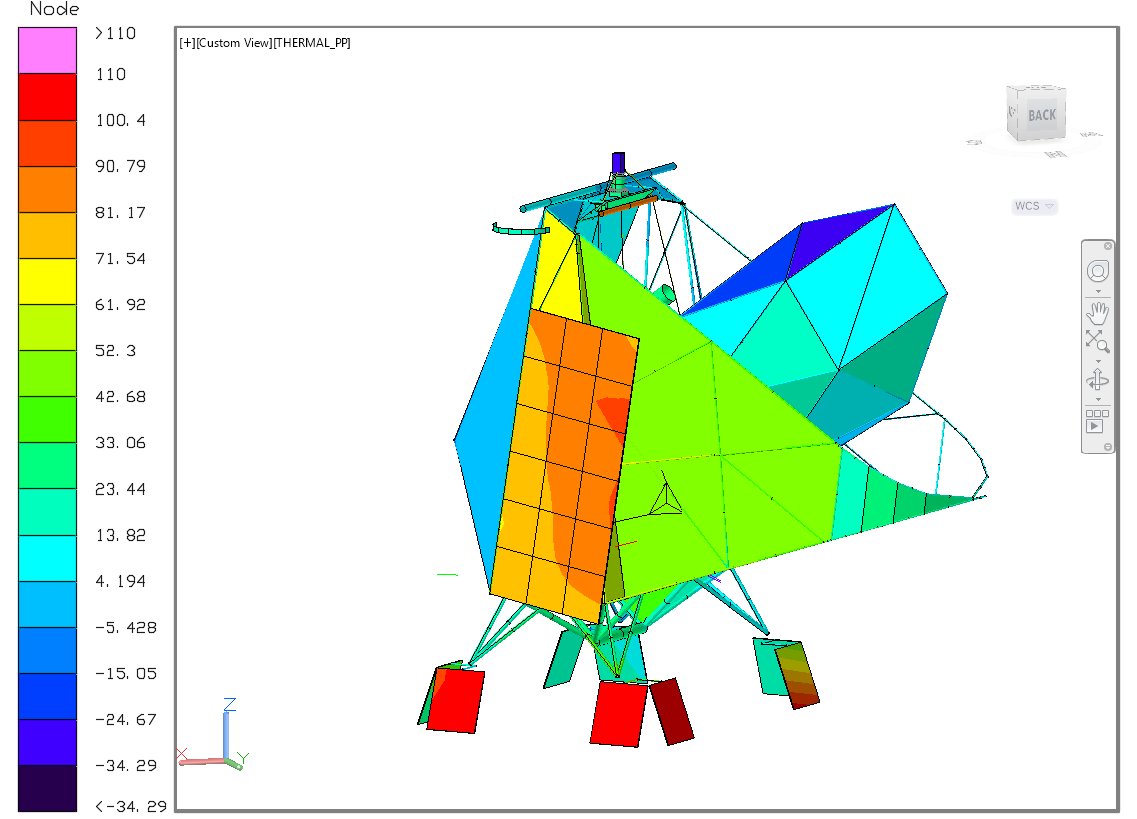}
    \\
    \includegraphics[width=0.45\textwidth]{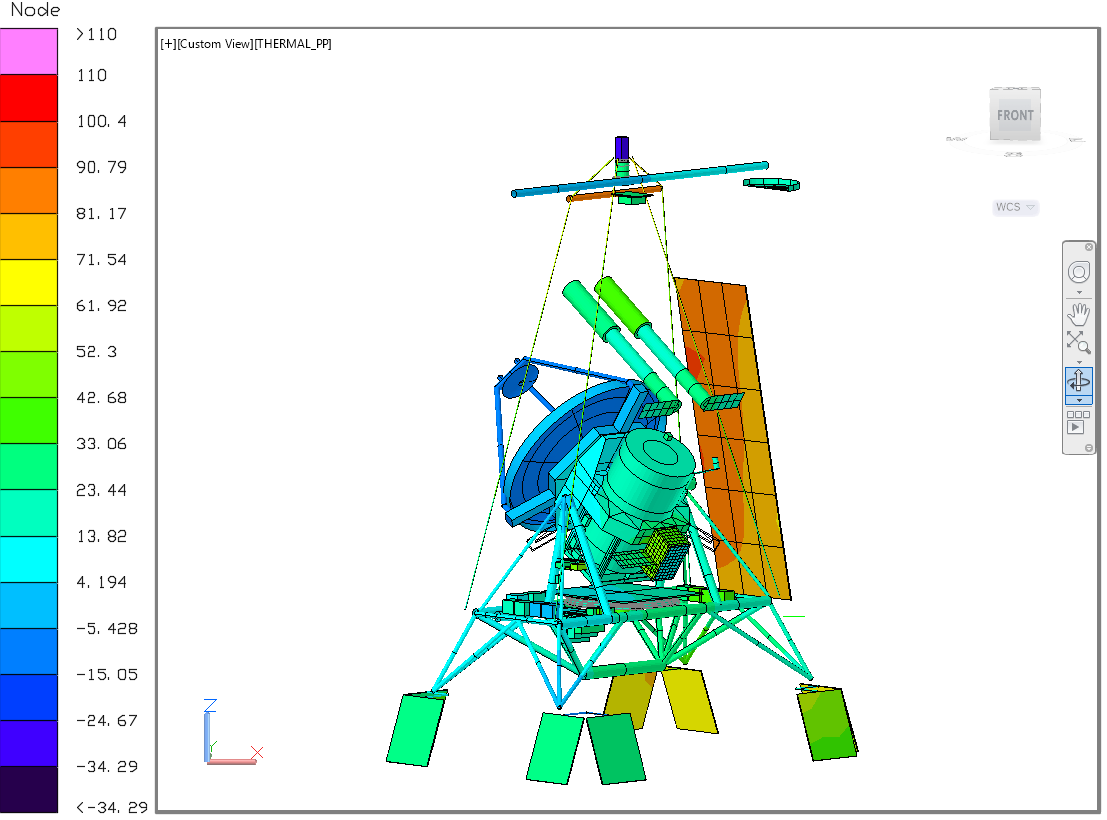}
    \caption{Final temperature results from a thermal desktop simulation of a flight pointing. The starboard side of the payload is significantly warmed by the solar radiation while the mylar shielding allows the instrument to remain between -5 and 43 C.}
    \label{fig:thermal_sims}
\end{figure*}

Thermal dissipation and management are extremely important in the near-vacuum environment of a stratospheric balloon. In addition to the power consumed and dissipated by the instrument itself, one must take into account the heat load from the Sun, the cooling from space, the reflected load from the surface of the Earth, and the vacuum environment with zero convective heat transfer. To study the temperature distribution, the BLAST-TNG team used a strategy of iteratively modeling and testing different components to verify that they would maintain a temperature within their operating limits. The model for the BLAST-TNG payload was simulated in Thermal Desktop (International Headquarters C$\&$R Technologies, Inc., Boulder, CO, USA), see figure \ref{fig:thermal_sims}, which employs a heat transfer and fluid dynamics differential equation solution method to model the thermal properties of the CAD drawing throughout an arbitrary period of time. 

The thermal model was built from power dissipation specifications of the various components on the gondola, as well as thermal conductivities based on materials and mounting interfaces. Since the largest source of uncertainties in the simulation results is improperly modeling the thermal interfaces, we tested our hardware interfaces and fed the results back into the simulation to refine our model. 

Tests were performed in both vacuum chambers for simple components or in thermal vacuum chambers for more complex components such as the ROACH readout system. Tests for small and simple components that require only a vacuum chamber tests were performed at the University of Pennsylvania. Tests that required the use of a thermal vacuum chamber were performed during the integration campaign in 2018 in Palestine, TX at the CSBF facility.

\begin{figure}
    \centering
    \includegraphics[scale=0.6]{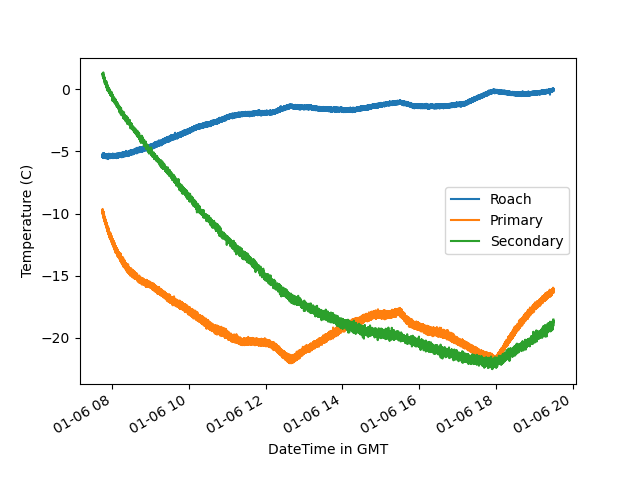}
    \caption{Temperature throughout the flight of some components}
    \label{fig:temps}
\end{figure}

Temperatures measured in flight are presented in figure \ref{fig:temps}. As shown in the plot, the temperatures of both mirrors were close to the predicted ones from the simulation. Also the enclosure of the readout system is relatively cold, giving a good indication that the thermal management of this component worked very well. Indeed, this is the major source of heat on the payload and this temperature gives a good indication that the roaches are well heat sunk to the gondola to dissipate the heat.

\section{Optical Performance}
\label{sec:optics}

\begin{figure}
\centering
\includegraphics[scale=0.8]{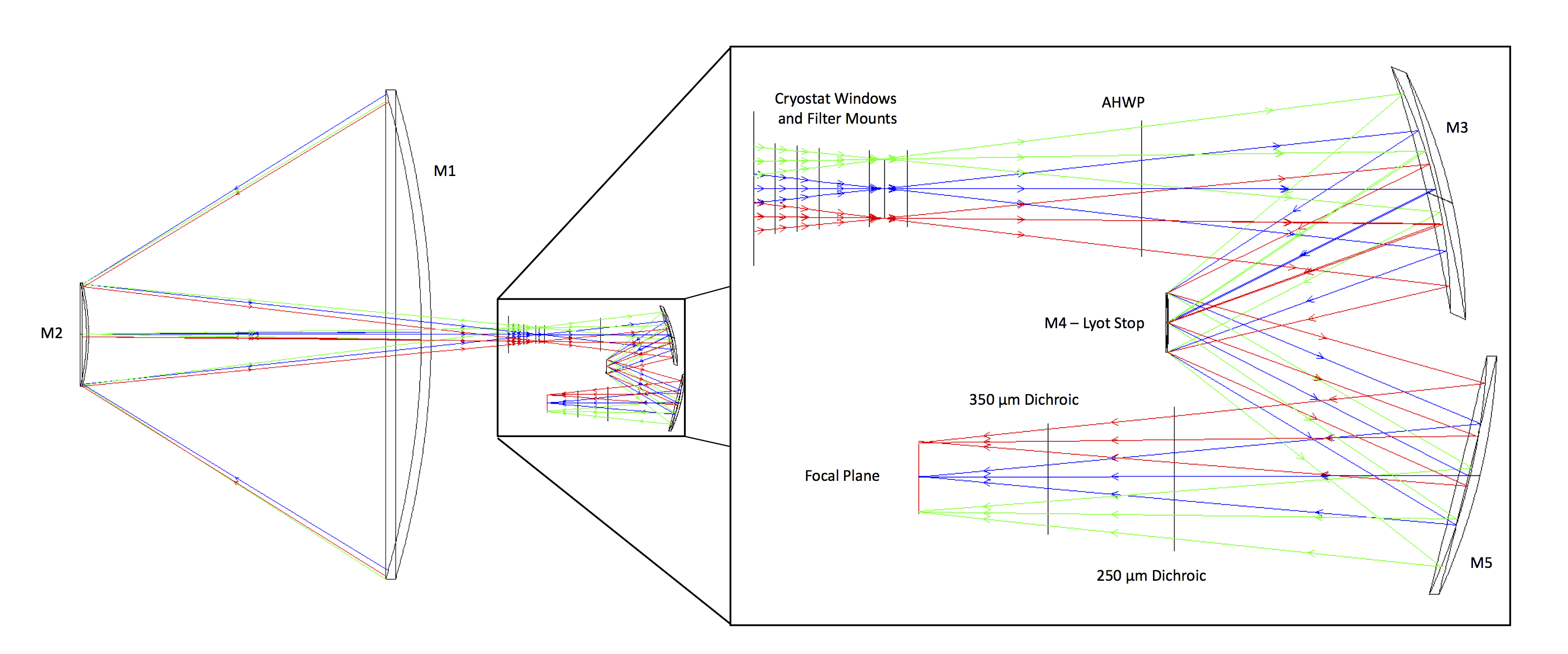}
\caption{Ray trace of the BLAST-TNG optics design from Zemax design software. The left-hand side shows the on-axis Cassegrain telescope formed by the primary (M1) and secondary mirrors (M2). The Cassegrain focus lies within the 4 K optics box, shown in the small rectangle. The light is refocused on the array with a modified Offner relay formed by the three mirrors M3, M4 (the Lyot stop), and M5. The locations of the three focal plane arrays are shown schematically.} 
\label{figure:optics_design}
\end{figure}

The BLAST-TNG optics design was based on a 2.5 m aperture on-axis Cassegrain telescope, with a carbon fiber composite primary mirror and an aluminum secondary mirror. The secondary mirror was mounted on three linear actuators which can move the secondary in piston/tip/tilt to account for changes in telescope focus due to differential thermal contraction of the telescope mirrors, CFRP support struts, and the aluminum gondola. The cryogenic reimaging optics are composed of a modified Offner relay which affords stability and ease of control over the beam illumination via a single-plane design on an optical bench. The optical design is shown in Fig. \ref{figure:optics_design}. The BLAST-TNG telescope was designed to operate over a wide field of view, with the high resolution necessary to resolve 0.1 pc scale structures in nearby molecular clouds. With its 23\arcmin FoV, the footprint of the detectors are offset over the 2.5 m aperture of the primary mirror. To reduce the risk of spillover, the cryogenic Lyot stop tapers the illumination of the primary mirror to 2.33 m (for the central beam). 

The telescope was developed through a commercial partnership with Alliance SpaceSystems, LLC (formerly Vanguard Space Technologies), through a NASA Small Business and Innovation Research grant (SBIR). The design was comprised of three major components: the primary and secondary mirrors which form the telescope beam, and a CFRP optical bench to which the primary and secondary mirror components mount. 

\begin{table}
\centering
\begin{tabular}{cll}
    \toprule
    Element & Parameter & Value \\
    %\colnumbers
    \midrule
    \multirow{3}[2]{*}{Primary Mirror} & Clear Aperture & \multicolumn{1}{r}{\diameter 2500.0 mm} \\
              & Radius of Curvature & \multicolumn{1}{r}{4132.11 mm} \\
              & Conic Constant & \multicolumn{1}{r}{-1.000000} \\
    \midrule
    \multirow{3}[2]{*}{Secondary Mirror} & Clear Aperture & \multicolumn{1}{r}{\diameter 573.0 mm} \\
              & Radius of Curvature & \multicolumn{1}{r}{1209.83 mm} \\
              & Conic Constant & \multicolumn{1}{r}{-2.380136} \\
    \midrule
    \multirow{6}[2]{*}{Telescope} & \multicolumn{1}{l}{Primary Vertex to Secondary Vertex} & \multicolumn{1}{r}{1590.0 mm} \\
              & EFL   & \multicolumn{1}{r}{9698.82 mm} \\
              & F/\#  & \multicolumn{1}{r}{3.87953} \\
              & FOV   & \multicolumn{1}{r}{23\arcmin} \\
              & Obscuration Ratio & \multicolumn{1}{r}{7.871\%} \\
              & Strut Obscuration Ratio & \multicolumn{1}{r}{2.618\%} \\
    \bottomrule
\end{tabular}
\caption{BLAST-TNG Telescope Optical Prescription}
\label{table:optical_prescription}
\end{table}

Before the 2020 flight extensive tests were performed on the ground to verify the surface accuracy of the two mirrors. Metrology on the components was performed using two FARO (FARO Global Headquarters, 250 Technology Park Lake Mary, FL 32746, USA) technologies, the laser tracker for large scale/object measurements and an arm for smaller/more confined measurements. The primary mirror mold was measured to have an 8.73\um RMS error relative to the design, slightly above the desired errors but within acceptable limits. Instead, for the secondary mirror, the surface error was measured to be less than 2\um rms in the area measured with a profilometer.

\begin{table}[h]
    \centering
    \begin{tabular}{cc}
       \toprule
       Parameter  &  Value \\
       \midrule
       Telescope total WFE  & $\leq$10 $\mu$m RMS \\
       \midrule
       \multirow{3}[0]{*}{Primary Mirror SE } &  $\leq$ 7 \um \space RMS on 50-250 cm scales \\
          & $\leq$ 4 \um \space RMS on 5-50 cm scales \\
          & $\leq$ 2 \um \space RMS on 0-5 cm scales \\
    \midrule 
    Secondary Mirror SE & $\leq$ 1 \um RMS \\
    \bottomrule
    \end{tabular}
    \caption{Target surface and wavefront errors for BLAST optics.}
    \label{tab:WFEs}
\end{table}

Before the flight, we aligned and focused the primary-secondary mirror system. These measurements were done using the FARO laser tracker after the telescope was installed on the gondola inner frame in December 2018. Given the possibility of using actuators to modify the secondary in-flight, the system only needed to be roughly focused before launch to prepare for focusing during initial calibration and checkout scans.

During the flight, the secondary mirror was adjusted in tip/tilt/piston via three stepper-motor based linear actuators each fitted with optical encoders (UltraMotion, 22355 CR 48, $\#$ 21 Cutchogue, NY 11935, D-A.083-AP702E5DIFF-1-/4-LT), which allow nearly 1 \um positioning resolution and have a 25 mm stroke length.

\begin{figure}
    \centering
    \includegraphics[width=\textwidth]{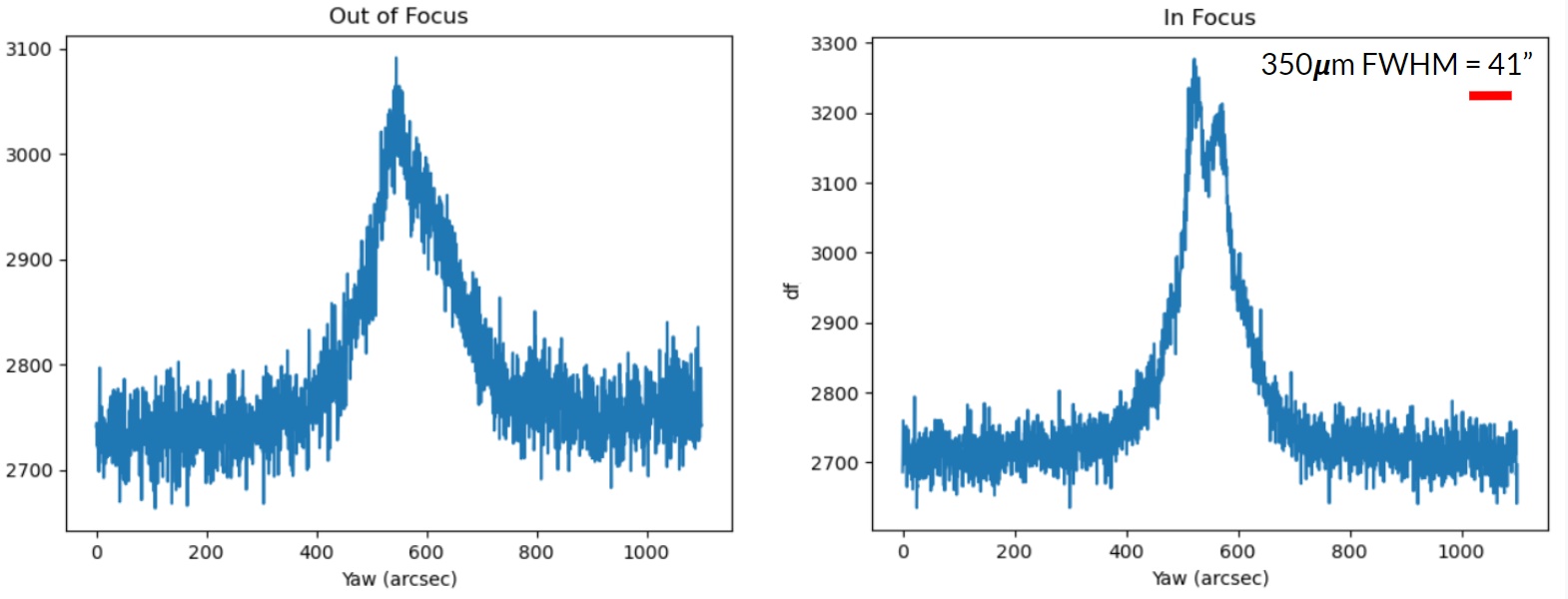}
    \caption{Azimuthal scans across the double-peaked Galactic source RCW 92B before and after focusing the telescope are shown respectively in the left and right panels.}
    \label{fig:focusing}
\end{figure}

Results obtained during in-flight focusing on the double-peaked Galactic HII region RCW 92B are presented in Figure~\ref{fig:focusing}. The traces shown are signals measured by a single detector in the 350$\mu$m array during a single pass across the target before (left) and after (right) focusing.  The separation in cross-elevation of the two components of RCW 92B is about 50 arcsec, which is close to our expected angular resolution of 41 arcsec (FWHM) for this waveband.  Note that the two sources are indistinguishable in the scan made prior to focusing, but are separated in the scan made after moving the secondary mirror actuators.

\section{Conclusions}

The BLAST program has already shown the importance of the study of the submm sky from a balloon platform with great scientific results\cite{fissel_blastpol}$^,$\cite{Soler_2017}$^,$\cite{Fissel_2019}. Unfortunately, this flight will not be able to provide that scientific outcome that we were expecting. However, BLAST-TNG provided invaluable information from a technology point of view. As shown in the previous sections, every subsystem worked as expected and when we had problems during flight, e.g. PMCs contaminating the star cameras images, we were able to solve those issues after flight. These subsystems are the baseline for the next iteration of the BLAST program, BLAST Observatory, that has just been proposed to NASA. Finally, BLAST-TNG was also the first kilopixel KIDs instrument to fly, and the gondola platform was able to satisfy all the requirements optically and in terms of power consumption for such large focal plane.

% References
\bibliography{References}

\begin{thebibliography}{10}

\bibitem{planck_XIX}
{Planck Collaboration}, {Ade}, P.~A.~R., {Aghanim}, N., {Alina}, D., {Alves},
  M.~I.~R., {Armitage-Caplan}, C., {Arnaud}, M., {Arzoumanian}, D., {Ashdown},
  M., {Atrio-Barandela}, F., and et~al., ``{Planck intermediate results. XIX.
  An overview of the polarized thermal emission from Galactic dust},'' {\em
  \aap}~{\bf 576},  A104 (Apr. 2015).

\bibitem{alma}
{Hull}, C.~L.~H., {Girart}, J.~M., {Tychoniec}, {\L}., {Rao}, R., {Cort{\'e}s},
  P.~C., {Pokhrel}, R., {Zhang}, Q., {Houde}, M., {Dunham}, M.~M.,
  {Kristensen}, L.~E., {Lai}, S.-P., {Li}, Z.-Y., and {Plambeck}, R.~L.,
  ``{ALMA Observations of Dust Polarization and Molecular Line Emission from
  the Class 0 Protostellar Source Serpens SMM1},'' {\em \apj}~{\bf 847},  92
  (Oct. 2017).

\bibitem{sofia}
Bittner, H. et~al., ``{SOFIA} primary mirror assembly: Structural properties
  and optical performance,'' {\em Proc. SPIE}~{\bf 4857} (2003).

\bibitem{pol2_jcmt}
{Friberg}, P., {Bastien}, P., {Berry}, D., {Savini}, G., {Graves}, S.~F., and
  {Pattle}, K., ``{POL-2: a polarimeter for the James-Clerk-Maxwell
  telescope},'' in [{\em Millimeter, Submillimeter, and Far-Infrared Detectors
  and Instrumentation for Astronomy VIII}{\nolinebreak\hspace{0.1em}]},  {\em
  Proc. SPIE} {\bf 9914},  991403 (July 2016).

\bibitem{ccat_prime_parshley}
{Parshley}, S.~C., {Kronshage}, J., {Blair}, J., {Herter}, T., {Nolta}, M.,
  {Stacey}, G.~J., {Bazarko}, A., {Bertoldi}, F., {Bustos}, R., {Campbell},
  D.~B., {Chapman}, S., {Cothard}, N., {Devlin}, M., {Erler}, J., {Fich}, M.,
  {Gallardo}, P.~A., {Giovanelli}, R., {Graf}, U., {Gramke}, S., {Haynes},
  M.~P., {Hills}, R., {Limon}, M., {Mangum}, J.~G., {McMahon}, J., {Niemack},
  M.~D., {Nikola}, T., {Omlor}, M., {Riechers}, D.~A., {Steeger}, K.,
  {Stutzki}, J., and {Vavagiakis}, E.~M., ``{CCAT-prime: a novel telescope for
  sub-millimeter astronomy},'' in [{\em Ground-based and Airborne Telescopes
  VII}{\nolinebreak\hspace{0.1em}]},  {\em Society of Photo-Optical
  Instrumentation Engineers (SPIE) Conference Series} {\bf 10700},  107005X
  (Jul 2018).

\bibitem{houde_2009_turbulence}
{Houde}, M., {Vaillancourt}, J.~E., {Hildebrand}, R.~H., {Chitsazzadeh}, S.,
  and {Kirby}, L., ``{Dispersion of Magnetic Fields in Molecular Clouds.
  II.},'' {\em \apj}~{\bf 706},  1504--1516 (Dec. 2009).

\bibitem{soler_hennebelle}
{Soler}, J.~D. and {Hennebelle}, P., ``{What are we learning from the relative
  orientation between density structures and the magnetic field in molecular
  clouds?},'' {\em \aap}~{\bf 607},  A2 (Oct 2017).

\bibitem{soler_sims_2013}
{Soler}, J.~D., {Hennebelle}, P., {Martin}, P.~G., {Miville-Desch{\^e}nes},
  M.-A., {Netterfield}, C.~B., and {Fissel}, L.~M., ``{An Imprint of Molecular
  Cloud Magnetization in the Morphology of the Dust Polarized Emission},'' {\em
  \apj}~{\bf 774},  128 (Sept. 2013).

\bibitem{jow_hro}
{Jow}, D.~L., {Hill}, R., {Scott}, D., {Soler}, J.~D., {Martin}, P.~G.,
  {Devlin}, M.~J., {Fissel}, L.~M., and {Poidevin}, F., ``{An application of an
  optimal statistic for characterizing relative orientations},'' {\em
  \mnras}~{\bf 474},  1018--1027 (Feb 2018).

\bibitem{fissel_blastpol}
{Fissel}, L.~M., {Ade}, P.~A.~R., {Angil{\`e}}, F.~E., {Ashton}, P., {Benton},
  S.~J., {Devlin}, M.~J., {Dober}, B., {Fukui}, Y., {Galitzki}, N., {Gandilo},
  N.~N., {Klein}, J., {Korotkov}, A.~L., {Li}, Z.-Y., {Martin}, P.~G.,
  {Matthews}, T.~G., {Moncelsi}, L., {Nakamura}, F., {Netterfield}, C.~B.,
  {Novak}, G., {Pascale}, E., {Poidevin}, F., {Santos}, F.~P., {Savini}, G.,
  {Scott}, D., {Shariff}, J.~A., {Diego Soler}, J., {Thomas}, N.~E., {Tucker},
  C.~E., {Tucker}, G.~S., and {Ward-Thompson}, D., ``{Balloon-Borne
  Submillimeter Polarimetry of the Vela C Molecular Cloud: Systematic
  Dependence of Polarization Fraction on Column Density and Local
  Polarization-Angle Dispersion},'' {\em \apj}~{\bf 824},  134 (June 2016).

\bibitem{draine_fraisse_2009}
{Draine}, B.~T. and {Fraisse}, A.~A., ``{Polarized Far-Infrared and
  Submillimeter Emission from Interstellar Dust},'' {\em \apj}~{\bf 696},
  1--11 (May 2009).

\bibitem{draine_hensley}
{Draine}, B.~T. and {Hensley}, B., ``{Magnetic Nanoparticles in the
  Interstellar Medium: Emission Spectrum and Polarization},'' {\em \apj}~{\bf
  765},  159 (Mar. 2013).

\bibitem{Lowe2020}
Lowe, I., Ade, P., Ashton, P., Austermann, J., Coppi, G., Cox, Erin~Devlin, M.,
  Dober, B., Fissel, L., Galitzki, N., Gao, J., Gordon, S., Groppi, C., Hilton,
  G., Hubmayr, J., Klein, J., Li, D., Lourie, N., Manii, H., Mauskopfi, P.,
  McKenneyd, C., Nati, F., Novak, G., Pisano, G., Romualdez, J., Sinclair, A.,
  Tucker, C., Ullom, J., Vissers, M., Wheeler, C., and Williams, P.,
  ``{Characterization, deployment, and in-flight performance of the BLAST-TNG
  cryogenic receiver},'' {\em Proc. SPIE} (2020).

\bibitem{chapman_stars}
{Chapman}, D., {Didier}, J., {Hanany}, S., {Hillbrand}, S., {Limon}, M.,
  {Miller}, A., {Reichborn-Kjennerud}, B., {Tucker}, G., and {Vinokurov}, Y.,
  ``{STARS: a software application for the EBEX autonomous daytime star
  cameras},'' in [{\em Software and Cyberinfrastructure for Astronomy
  III}{\nolinebreak\hspace{0.1em}]},  {\em Proc. SPIE} {\bf 9152},  915212
  (July 2014).

\bibitem{astrometry_net}
Lang, D., Hogg, D.~W., Mierle, K., Blanton, M., and Roweis, S.,
  ``Astrometry.net: Blind astrometric calibration of arbitrary astronomical
  images,'' {\em The Astronomical Journal}~{\bf 139},  1782–1800 (Mar 2010).

\bibitem{Soler_2017}
Soler, J.~D., Ade, P. A.~R., Angilè, F.~E., Ashton, P., Benton, S.~J., Devlin,
  M.~J., Dober, B., Fissel, L.~M., Fukui, Y., Galitzki, N., and et~al., ``The
  relation between the column density structures and the magnetic field
  orientation in the vela c molecular complex,'' {\em Astronomy \&
  Astrophysics}~{\bf 603},  A64 (Jul 2017).

\bibitem{Fissel_2019}
Fissel, L.~M., Ade, P. A.~R., Angilè, F.~E., Ashton, P., Benton, S.~J., Chen,
  C.-Y., Cunningham, M., Devlin, M.~J., Dober, B., Friesen, R., and et~al.,
  ``Relative alignment between the magnetic field and molecular gas structure
  in the vela c giant molecular cloud using low- and high-density tracers,''
  {\em The Astrophysical Journal}~{\bf 878},  110 (Jun 2019).

\end{thebibliography}
\bibliographystyle{spiebib} % makes bibtex use spiebib.bst

\end{document}